\documentclass[aps,prl,twocolumn,showpacs,amsmath,amssymb,nofootinbib, preprintnumbers]{revtex4-1}
\usepackage{graphicx}
\usepackage{float}
\usepackage{color}

\newcommand{\be}{\begin{equation}}
\newcommand{\ee}{\end{equation}}
\newcommand{\ba}{\begin{eqnarray}}
\newcommand{\ea}{\end{eqnarray}}

\newcommand{\beq}{\begin{equation}}
\newcommand{\eeq}{\end{equation}}
\newcommand{\beqa}{\begin{eqnarray}}
\newcommand{\eeqa}{\end{eqnarray}}

\begin{document}

\title{Super-Entropic Black Holes}

\author{Robie A. Hennigar}
\email{rhennigar@uwaterloo.ca}
\affiliation{Department of Physics and Astronomy, University of Waterloo,
Waterloo, Ontario, Canada, N2L 3G1}

\author{David Kubiz\v n\'ak}
\email{dkubiznak@perimeterinstitute.ca}
\affiliation{Perimeter Institute, 31 Caroline St. N., Waterloo,
Ontario, N2L 2Y5, Canada}
\affiliation{Department of Physics and Astronomy, University of Waterloo,
Waterloo, Ontario, Canada, N2L 3G1}

\author{Robert B. Mann}
\email{rbmann@sciborg.uwaterloo.ca}
\affiliation{Department of Physics and Astronomy, University of Waterloo,
Waterloo, Ontario, Canada, N2L 3G1}

%
\date{November 17, 2014}

\begin{abstract}
We construct a new class of rotating AdS black hole solutions with non-compact event horizons of finite area in
any dimension and study their thermodynamics. 
In four dimensions these black holes are solutions to gauged supergravity.   
We find that their entropy exceeds the maximum implied from the conjectured Reverse Isoperimetric Inequality, which states that for a given 
thermodynamic volume, the black hole entropy is maximized for Schwarzschild-AdS.  We 
use this result to suggest more stringent conditions under which this conjecture may hold.
\end{abstract}






\maketitle



The study of the thermodynamics of black holes in anti de Sitter space (AdS) has received much attention since the seminal paper of Hawking and Page~\cite{Hawking:1982dh}.  The AdS case is of particular  interest  because thermodynamic equilibrium is straightforwardly defined and because they admit a gauge duality description via a dual thermal field theory. 

A novel topic of active study in recent years 
is the proposal that the mass of an AdS black hole should be interpreted as the enthalpy of  spacetime \cite{Kastor:2009wy}.  This idea is a consequence of considering the cosmological constant $\Lambda$ to be a thermodynamic variable 
\cite{Creighton:1995au} analogous to pressure in the first law 
\cite{Caldarelli:1999xj, Kastor:2009wy, Dolan:2010ha, Dolan:2011xt, Dolan:2014jva, Cvetic:2010jb, Kubiznak:2012wp, Altamirano:2014tva, Johnson:2014yja, Johnson:2014xza, MacDonald:2014zaa}, 
%
which has a number of implications. First, as a thermodynamic quantity, $\Lambda$ must 
have a conjugate variable, the natural interpretation of which is a {\em thermodynamic volume} associated with the black hole.\footnote{{For a discussion on extending  thermodynamic volume beyond black hole spacetimes, see \cite{Johnson:2014xza, MacDonald:2014zaa}.}} Second, in the  presence of a  nonzero $\Lambda$, 
the standard {\em Smarr formula} no longer holds.
This problem is remedied when $\Lambda$
is permitted to vary in the first law and the corresponding term is added to the Smarr relation \cite{Kastor:2009wy}. 
Third, the extended phase space allows one to rewrite black hole thermodynamic equations as {\em equations of state} analogous to those of 
everyday simple substances, obtaining, for example, a gravitational analogue of the Van der Waals fluid, 
triple points and reentrant phase transitions (for a  review see \cite{Altamirano:2014tva})
and the notion of a holographic heat engine
\cite{Johnson:2014yja}.

The proposed relationship between the cosmological constant and the pressure is
\be\label{eq:pressure}
P = -\frac{1}{8\pi}\Lambda = \frac{(d-1)(d-2)}{16\pi l^2}\,,
\ee
where $d$ is the number of spacetime dimensions.  
The thermodynamic volume $V$ for the asymptotically AdS black hole spacetimes is then defined so that the following
extended first law of black hole thermodynamics holds:
\be\label{eq:first-law}
\delta M = T\delta S +\sum_i \Omega_i\delta J_i + \Phi\delta Q + V \delta P\,,
\ee
a result supported by geometric arguments \cite{Kastor:2009wy}.
Here, $M, J, T$  and $S$ stand for the mass, angular momentum, temperature, and the entropy of the black hole, while the $\Omega_i$ are the 
angular velocities and $\Phi$ is the 
electric potential, 
all measured with respect to infinity.
The corresponding Smarr relation 
\be\label{Smarr}
\frac{d-3}{d-2}M=TS+\sum_i \Omega_i J_i+\frac{d-3}{d-2}\Phi Q-\frac{2}{d-2}VP\,,
\ee
can be derived from a scaling (dimensional) argument~\cite{Kastor:2009wy}.

An interesting property of the thermodynamic volume is that, in all cases studied so far, it satisfies what is known as the {\em Reverse Isoperimetric Inequality}~\cite{Cvetic:2010jb, Altamirano:2014tva}.   Indeed, it was conjectured in~\cite{Cvetic:2010jb} 
that the isoperimetric ratio 
\be\label{eq:ipe-ratio}
\mathcal{R}=\left(\frac{(d-1) {V}}{\omega_{d-2}}\right)^{\frac{1}{d-1}}\left(\frac{\omega_{d-2}}{ {A}}\right)^{\frac{1}{d-2}}
\ee
always satisfies $\mathcal{R} \ge 1$. 
Here ${V}$ is the thermodynamic volume, ${A}$ is the horizon area, and $\omega_d$ stands for the area of 
the space orthogonal to constant $(t,r)$ surfaces; for 
a $d$-dimensional unit sphere,
$\omega_d = \frac{2\pi^{\frac{d+1}{2}}}{\Gamma\left(\frac{d+1}{2}\right)}$.
This result  can be interpreted as  implying that Schwarzschild AdS black holes are `maximally entropic':  for a black hole of a given ``volume'' $V$ its entropy is maximized for Schwarzschild-AdS. 

Here we construct a new ultraspinning limit to the singly-spinning Kerr-AdS metric in $d$ space-time dimensions that yields a new class of black hole solutions whose entropies exceed this maximum bound.  When $d=4$ these metrics are equivalent (upon inclusion of charge) to a class of black hole solutions of   gauged supergravity in 4 dimensions, recently derived in~\cite{Gnecchi:2013mja}  and later elaborated upon in~\cite{Klemm:2014rda}.  Insofar as the solutions we construct are understood as string theory ground states,  through the AdS/CFT correspondence, topics such as microscopic degeneracy can be studied~\cite{Strominger:1996sh}.  The class of  black holes we construct all have  horizons  that are noncompact yet have finite area (and therefore entropy). We find that this particular feature is sufficient to ensure that  their entropy exceeds the maximal bound implied by the
 Reverse Isoperimetric Inequality \cite{Cvetic:2010jb}; as such they provide the first counterexample to this conjecture. 
 Such black holes are therefore `{\em super-entropic}'.

Let us start with the 4-dimensional Kerr--Newman-AdS solution \cite{Carter:1968ks}, written in the `standard Boyer--Lindquist form' \cite{Hawking:1998kw}
\ba\label{KNADS}
ds^2&=&-\frac{\Delta_a}{\Sigma_a}\left[dt-\frac{a\sin^2\!\theta}{\Xi}d\phi\right]^2
+\frac{\Sigma_a}{\Delta_a} dr^2+\frac{\Sigma_a}{S}d\theta^2\nonumber\\
&&+\frac{S\sin^2\!\theta}{\Sigma_a}\left[a dt-\frac{r^2+a^2}{\Xi}d\phi\right]^2\,,\nonumber\\
{\cal A}&=&-\frac{qr}{\Sigma_a}\left(dt-\frac{a\sin^2\!\theta}{\Xi}d\phi\right)\,,
\ea
where
\ba
\Sigma_a&=&r^2+a^2\cos^2\!\theta\,,\quad \Xi=1-\frac{a^2}{l^2}\,,
\quad S=1-\frac{a^2}{l^2}\cos^2\!\theta\,,\nonumber\\
\Delta_a&=&(r^2+a^2)\Bigl(1+\frac{r^2}{l^2}\Bigr)-2mr+q^2
\ea
with the horizon $r_h$ defined by $\Delta_a(r_h)=0$.
The thermodynamic quantities obeying \eqref{eq:first-law} were calculated in \cite{Caldarelli:1999xj, Cvetic:2010jb, Dolan:2011xt}; 
in particular, the thermodynamic volume $V$ and the horizon area $A$ read 
\ba
V&=&\frac{2\pi}{3}\frac{(r_h^2+a^2)(2r_h^2l^2+a^2l^2-r_h^2a^2)+l^2q^2a^2}{l^2\Xi^2r_h}\,,\nonumber\\
A&=&\frac{4\pi(r_h^2+a^2)}{\Xi}\,,
\ea
and satisfy the isoperimetric inequality $\mathcal{R}\geq 1$. 

Let us now consider a new ultraspinning limit  as follows. We first replace everywhere $\psi = \phi/\Xi$, and then take the
$a\to l$ limit. In this way we obtain a new solution of the Einstein--Maxwell equations, the super-entropic black hole, given by
\ba\label{KNADS2}
ds^2&=&-\frac{\Delta}{\Sigma}\left[dt-l\sin^2\!\theta d\psi\right]^2
+\frac{\Sigma}{\Delta} dr^2+\frac{\Sigma}{\sin^2\!\theta}d\theta^2\nonumber\\
&&+\frac{\sin^4\!\theta}{\Sigma}\left[l dt-(r^2+l^2)d\psi\right]^2\,,\nonumber\\
{\cal A}&=&-\frac{qr}{\Sigma}\left(dt-l\sin^2\!\theta d\psi\right)\,,
\ea
where
\ba
\Sigma=r^2+l^2\cos^2\!\theta\,,\quad \Delta=\Bigl(l+\frac{r^2}{l}\Bigr)^2-2mr+q^2\,.\quad
\ea
In this form, $\psi$ is a non-compact coordinate, which we now compactify 
\be\label{psiC}
\psi \sim \psi + \mu
\ee
where the parameter $\mu$  is dimensionless. 
The metric (\ref{KNADS2}) can be shown to be the same as that obtained in  
\cite{Gnecchi:2013mja, Klemm:2014rda} via the following change of coordinates:
\be
\tau=t\,,\quad p=l\cos\theta\,,\quad \sigma=-\psi/l\,,\quad L=\mu/l\,.
\ee 

The location of the horizon $r_+$ is
determined by the largest root of $\Delta(r)$.  In order for horizons to exist, the mass parameter must satisfy 
\be\label{eq:extremal-mass}
m \ge 2r_0\left(\frac{r_0^2}{l^2}+1\right)\,,\quad 
r^2_0 \equiv \frac{l^2}{3}\Bigl[-1+\left(4+\frac{3}{l^2}q^2\right)^{1/2}\Bigr]\,.
\ee
The case where equality holds 
corresponds to an extremal black hole. 

The fundamental thermodynamic parameters of the super-entropic black hole are 
\begin{align}
M&=\frac{\mu m}{2\pi}\,,\quad J=Ml\,,\quad 
\Omega = \frac{l}{r_+^2+l^2}\,,  \nonumber
\\
T&=\frac{1}{4\pi r_+}\left(3\frac{r_+^2}{l^2}-1-\frac{q^2}{l^2+r^2} \right), \nonumber \\ 
S&= \frac{\mu}{2}(l^2+r_+^2)=\frac{A}{4},\quad \Phi=\frac{qr_+}{r_+^2+l^2}, \;\;\; Q=\frac{\mu q}{2\pi}\, .
\label{eq:thermo_properties}
\end{align} 
The angular velocity $\Omega$  is that of the horizon and
the mass and angular momentum were computed using the method of conformal completion \cite{Ashtekar:1984zz, Ashtekar:1999jx, Das:2000cu} using the associated Killing vectors $\partial_\tau$ and $\partial_{\psi}$ respectively. Note also the `chirality condition' $J=Ml$.  

The thermodynamic 
characteristics displayed
here appear, at first glance, to differ from those presented in~\cite{Klemm:2014rda}.  However, the quantities are easily shown to be the same when one takes note of two points. First, the quantity $\mu$ in eq.~(\ref{eq:thermo_properties}) is related to $L$ from~\cite{Klemm:2014rda} 
by $\mu = lL$.  Second, the angular momentum computed in~\cite{Klemm:2014rda} is computed with respect to the coordinate $\sigma$, which has dimension $[L]^{-1}$, rendering it to be an angular momentum per unit length.  The quantities in (\ref{eq:thermo_properties}) all have scaling dimensions consistent with their Kerr-Newman-AdS counterparts.

We now consider the thermodynamics in extended phase space.  In addition to considering pressure and volume terms, we also consider $\mu$ as a thermodynamic parameter.   In the context of asymptotic Schr\"odinger geometries  the compactified null length can be interpreted as a chemical potential~\cite{Herzog:2008wg}.  As discussed in~\cite{Klemm:2014rda}, $\psi$ becomes identified as a compact null coordinate on the conformal boundary.  Since $\mu$ is associated with the compactification of $\psi$, we therefore interpret $\mu$ as being related to a chemical potential and denote its thermodynamic conjugate as $K$.

To determine $K$ and the thermodynamic volume $V$, we demand consistency of the extended first law:
\be\label{1stlaw}
dM = TdS +VdP + \Omega dJ + \Phi dQ + Kd\mu.
\ee
Doing so yields
\ba
V&=& \frac{r_+A}{3}=\frac{2}{3} \mu r_+ \left(r_+^2 + l^2 \right) \,,\label{eq:volume}\\
K &=& \frac{(l^2-r_+^2)\bigl[ (r_+^2+l^2)^2 + q^2 l^2\bigr]}{8\pi l^2 r_+(r_+^2 + l^2)}\,.\label{eq:Lconj}
\ea
The above thermodynamic quantities obey the Smarr relation \eqref{Smarr}; note that there is no contribution from a
$K\mu$-term as $\mu$ is a dimensionless quantity. 
It is interesting to note that the thermodynamic volume $V$ found here 
is reminiscent of the {\em naive geometric volume} (the integral of $\sqrt{-g}$  `inside' the event horizon) 
of the Kerr-AdS black hole (studied in detail in \cite{Cvetic:2010jb}), in
strong contrast to the traditional ultraspinning black holes, for which the naive geometric volume negligibly contributes to $V$ \cite{Altamirano:2014tva}.  Note that $V$ does not explicitly depend on the black hole charge $q$, 
as in the case of  non-rotating charged AdS black holes \cite{Kubiznak:2012wp}.

It is straightforward to see that these black holes are super-entropic.
Bearing  in mind that our space is compactified according to  \eqref{psiC}, the orthogonal 2-dimensional surface area takes the form 
$\omega_2=2\mu$.  Consequently, the isoperimetric ratio \eqref{eq:ipe-ratio} now reads
\be
\mathcal{R}=\left(\frac{r_+A}{2\mu}\right)^{1/3}\left(\frac{2\mu}{A}\right)^{1/2}
= \left(\frac{r_+^2}{r_+^2 + l^2}\right)^{1/6}< 1\,.
\ee
Hence we have shown that our black holes {\em always violate the Reverse Isoperimetric Inequality}.

This result stands in contrast to the `usual' ultraspinning limit of Kerr-AdS black holes \cite{Emparan:2003sy} in which, as $a \to l$, the isoperimetric ratio approaches infinity, maximally satisfying the reverse isoperimetric inequality.  The distinction arises because of the nature of the ultraspinning limit we are taking. Rather than keeping $M$  fixed and letting the horizon area approach zero as $a \to l$ \cite{Emparan:2003sy, Altamirano:2014tva}, here we require this limit be taken whilst demanding the horizon area remain finite.

Unfortunately this class of charged black holes does not have  interesting phase behaviour or critical phenomena.
For example, in the charge free case we obtain (using equations (\ref{eq:pressure}) and (\ref{eq:thermo_properties}) for the pressure and temperature, respectively)
\be
P=\frac{T}{v} + \frac{1}{2 \pi v^2}
\ee 
for the equation of state, where the specific volume, $v = 2r_+=6V/(A/l_{P}^2)$  \cite{Kubiznak:2012wp, Altamirano:2014tva}.  Since all terms on the right-hand side are positive, it is not possible for this black hole to exhibit critical behaviour in the charge-free case.   Inclusion of electric charge leads to a more complicated  equation of state, but likewise yields no interesting critical behaviour. The same  conclusion holds if  alternate definitions of the specific volume are employed (see, e.g. those in~\cite{Altamirano:2014tva}).

For completeness we connect our results with previous thermodynamic considerations \cite{Klemm:2014rda} for these black holes.  
The quantities $M$ and $J$ are not independent;   consequently we should consider $M = M(S,P, Q,\mu)$ in deriving the Smarr formula.  A more convenient choice of thermodynamic variables is $ {L}_\pm = \frac{1}{2}\left(M \pm J/l\right)$, where $L_-$ vanishes due to the chirality condition  $J= Ml$ \cite{Klemm:2014rda}. It is straightforward to show that the first law 
(\ref{1stlaw}) becomes
\ba
TdS &=& \bigg(1-\Omega \sqrt{\frac{3}{8 \pi P}}\bigg)d {L}_+  -
\bigg(V-\frac{\Omega L_+}{8 P}\sqrt{\frac{6}{\pi P}}\bigg)dP \nonumber\\ 
&-&K d\mu -\Phi dQ
\label{eq:newFirstLaw}
\ea
and scaling arguments imply that the Smarr formula is
\be\label{eq:newSmarr}
Z {L_+} = 2(TS - V^\prime P) + \Phi Q\,,
\ee
where $Z$ and $V^\prime$ are the respective thermodynamic conjugates to $L_{+}$  and $P$,  from eq.~(\ref{eq:newFirstLaw}).   Note that, in the case where $\Lambda$ and $\mu$ are not considered  thermodynamic quantities,
 the terms proportional to $dP$ and $d\mu$ are not present and the standard form of the first law \cite{Klemm:2014rda} is recovered.

The preceding considerations suggest that perhaps
\be
V^\prime = \Bigl(V-\frac{\Omega L_+}{8 P}\sqrt{\frac{6}{\pi P}}\Bigr)
\label{eq:volumePrime}
\ee
could be regarded as  the thermodynamic volume rather than $V$ from eq.~(\ref{eq:volume}).  However,
 unlike $V$,  $V^\prime$  suffers from the drawback of not being strictly positive for all values of the parameters, and the quantity $L_+$ in (\ref{eq:newFirstLaw}, \ref{eq:newSmarr}) does not correspond to either a mass or an angular momentum.  
Furthermore, making this identification
does not alter our basic result, as we find that the Reverse Isoperimetric Inequality is still violated using $V^\prime$.   For these reasons we contend that  $V$ defined by eq.~(\ref{eq:volume}) should be regarded as the thermodynamic volume for this class of black holes.

Based on the limiting procedure introduced earlier, it is a straightforward matter to generalize the metric (\ref{KNADS2})
to $d$ dimensions.  Starting with the singly spinning ($a_1=a$ and other $a_i=0$) Kerr-AdS solution, replacing everywhere $\phi = \psi \Xi$ and then taking the limit $a \to l$ we obtain  
\ba\label{Singlyspinning}
ds^2&=&-\frac{\Delta}{\rho^2}(dt-{l}\sin^2\!\theta d\psi)^2+
\frac{\rho^2}{\Delta}dr^2+\frac{\rho^2}{\sin^2\!\theta}d\theta^2\quad \nonumber\\
&+&
\frac{\sin^4\!\theta}{\rho^2}[ldt-({r^2\!+\!l^2})d\psi]^2\!+\!r^2\cos^2\!\theta d\Omega_{d-4}^2\,,\qquad\ \ 
\ea
where
\ba
\Delta&=&\Bigl(l+\frac{r^2}{l}\Bigr)^2-2mr^{5-d}\,,\ \rho^2=r^2+l^2\cos^2\!\theta\,,\quad
\ea
and $d\Omega_{d}^2$ denotes the metric element on a $d$-dimensional sphere.  In this form, $\psi$ is a noncompact coordinate, which we now compactify via $\psi \sim \psi + \mu$.  It is straightforward to show that the metric
\eqref{Singlyspinning} satisfies the Einstein-AdS equations. Setting $d=4$ we recover the metric (\ref{KNADS2})
with $q=0$.  Horizons exist in any dimension $d\geq 5$ provided $m>0$.
  
The solution inherits a closed conformal Killing--Yano 2-form from the Kerr-AdS metric, given by 
$h=db$, where 
\be
b=(l^2\cos^2\!\theta-r^2)dt-l(l^2\cos^2\!\theta-r^2\sin^2\!\theta)d\psi\,.
\ee 
This object, together with the explicit symmetries of the metric guarantee complete integrability of geodesic motion as 
well as separability of the Hamilton--Jacobi, Klein--Gordon, and Dirac equations in the black hole background; see 
\cite{Frolov:2008jr} for analogous results in the Kerr-AdS case.

Computing the thermodynamic quantities for this solution in extended phase space, we find
\begin{eqnarray}\label{singlespin}
M&=&\frac{\omega_{d-2}}{8\pi} \left(d-2 \right)m  \,, \quad
J=\frac{2}{d-2}Ml\,,\quad \Omega=\frac{l}{r_+^2+l^2}\,,\nonumber\\
T&=&\frac{1}{4\pi r_+ l^2}\Bigr[ (d-5)l^2 + r^2(d-1)\Bigr]\,,\nonumber\\
S&=&\frac{\omega_{d-2}}{4}(l^2+r_+^2) r_+^{d-4}=\frac{A}{4}\,,\quad
V=\frac{r_+A}{d-1}\,,
\end{eqnarray} 
where
\be
\omega_{d} = \frac{\mu \pi ^{\frac{d-1}{2}}}{\Gamma\left(\frac{d+1}{2}\right)}
\ee
is the volume of the $d$-dimensional  unit `sphere'.
Here $\Omega$ is the angular velocity of the horizon and $J$ and $M$ have been computed via the method of conformal completion as the conserved quantities associated with the $\partial_\psi$ and $\partial_t$ Killing vectors, respectively.   
 
By varying $\mu$ we get the following expression for its conjugate quantity $K$:
\be
K=\frac{1}{\mu}(M-TS-\Omega J)\,.
\ee
One can easily verify that these thermodynamic quantities satisfy the Smarr formula \eqref{Smarr}.
The thermodynamic volume obtained here satisfies the isoperimetric inequality ($\mathcal{R} \le 1$) in all dimensions, 
and so this class of black holes is also super-entropic. 


To summarize, we have constructed a class of black hole solutions to Einstein-AdS gravity that result from taking a new ultraspinning limit of Kerr-AdS black holes in $d$-dimensions.  These black holes are super-entropic insofar as their entropy is larger than the maximum allowed by the  Reverse Isoperimetric Inequality \cite{Cvetic:2010jb} (shown to be obeyed by all previously known
black hole solutions): they have a greater entropy than their thermodynamic volume would naively allow.

We attribute this behaviour to be a consequence of their finite-area but noncompact event horizons.  We posit that
 the Reverse Isoperimetric Inequality Conjecture might hold under more stringent conditions:
  for a black hole with thermodynamic volume $\mathcal{V}$ and with compact horizon of area $\mathcal{A}$,
the ratio~(\ref{eq:ipe-ratio}) satisfies $\mathcal{R} \ge 1$.  The proof of this conjecture remains an interesting open question for further study.

Ultraspinning limits similar to the type we have taken here can be applied to multiply spinning Kerr-AdS black holes and charged Kerr-AdS black holes in higher dimensions, yielding further new classes of solutions. We shall elaborate further on this topic in the  near future \cite{USFuture}.

\section*{Acknowledgments}
This research was supported in part by Perimeter Institute for Theoretical Physics and by the Natural Sciences and Engineering Research Council of Canada. Research at Perimeter Institute is supported by the Government of Canada through Industry Canada and by the Province of Ontario through the Ministry of Research and Innovation.



\providecommand{\href}[2]{#2}\begingroup\raggedright\endgroup

\end{document}